\documentclass[fleqn,usenatbib,useAMS]{mnras}

\usepackage[T1]{fontenc}
\usepackage{ae,aecompl}

\DeclareRobustCommand{\VAN}[3]{#2}
\let\VANthebibliography\thebibliography
\def\thebibliography{\DeclareRobustCommand{\VAN}[3]{##3}\VANthebibliography}

\usepackage[english]{babel}
\usepackage{graphicx}
\graphicspath{{./}{LEGACY/}}
\usepackage{floatflt}
\usepackage{color}
\usepackage{verbatim}
\usepackage{caption}
\usepackage{subcaption}
\usepackage{amsmath}
\usepackage{wrapfig}

\usepackage{newtxtext,newtxmath}

\title[Bent-tube like astropause model]{New unexpected flow patterns in the problem of the stellar wind interaction with the interstellar medium: stationary ideal-MHD solutions}

\author[Korolkov and Izmodenov]{Sergey Korolkov$^{1,2}$ and  Vladislav Izmodenov$^{1,2,3}$%
\\
\author[0000-0002-1748-0982]{Vladislav Izmodenov}
$^{1}$Space Research Institute (IKI) Russian Academy of Sciences, Moscow, Russia \\
$^{2}$Lomonosov Moscow State University, Center for fundamental and applied mathematics, Moscow, Russia \\
$^{3}$National Research University Higher School of Economics}

\begin{document}
\label{firstpage}
\pagerange{\pageref{firstpage}--\pageref{lastpage}}
\maketitle

\begin{abstract}
The astropause (heliopause for the Sun) is the tangential discontinuity separating the stellar wind from the interstellar plasma. The global shape of the heliopause is a matter of debates. Two types of the shape are under discussion: comet-like and tube-like. In the second type the two-jets oriented toward the stellar rotation axis are formed by the action of azimuthal component of the stellar magnetic field. 
We explore a simplified global astrosphere in which: (1) the surrounding and moving with respect to the star circumstellar medium is fully ionized,  (2) the interstellar magnetic field is neglected, (3) the radial component of the stellar magnetic field is neglected as compared with the azimuthal component, (4) the stellar wind outflow is spherically symmetric and supersonic.  
We present the results of numerical 3D MHD modelling and explore how the global structure depends on the gas-dynamic Mach number of the interstellar flow, $M_\infty$, and the Alfvenic Mach number in the stellar wind.   
It is shown that  the astropause has a tube-like shape for small values of $M_\infty$.  The wings of the tube are distorted toward the tail as larger as larger the Mach number is. The new (to our knowledge) result is the reverse interstellar flow in the vicinity of  the astropause in the tail. The larger the interstellar Mach number is the narrower the reverse flow is. At some values of the Mach number the stellar wind overcomes the reverse interstellar flow and moves out in downwind. In this regime the astropause changes its topology from tube-like to sheet-like. 
\end{abstract}





\section{Introduction} \label{sec:intro}

\begin{figure*}
	\includegraphics[width=0.33\textwidth]{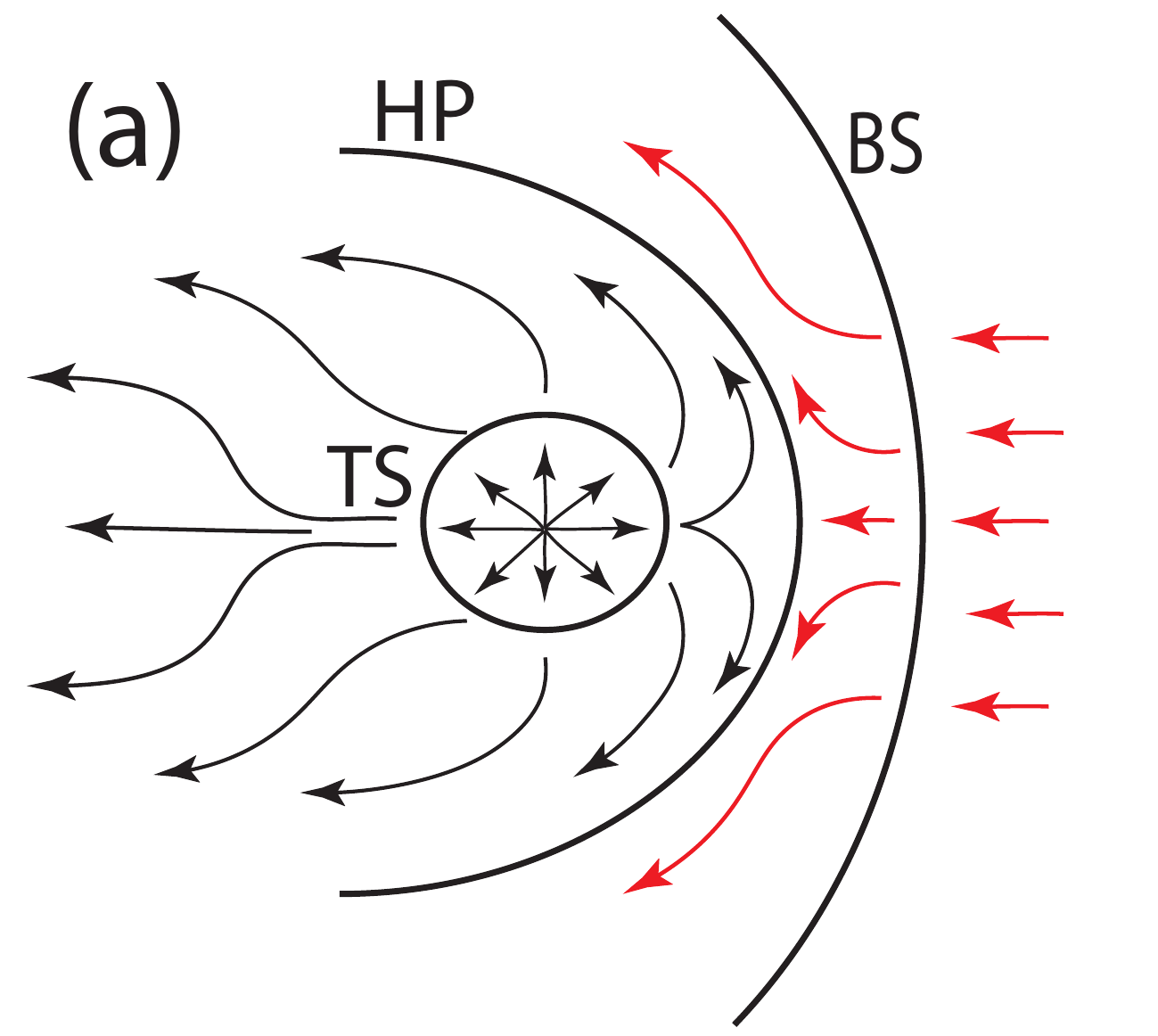}
	\includegraphics[width=0.33\textwidth]{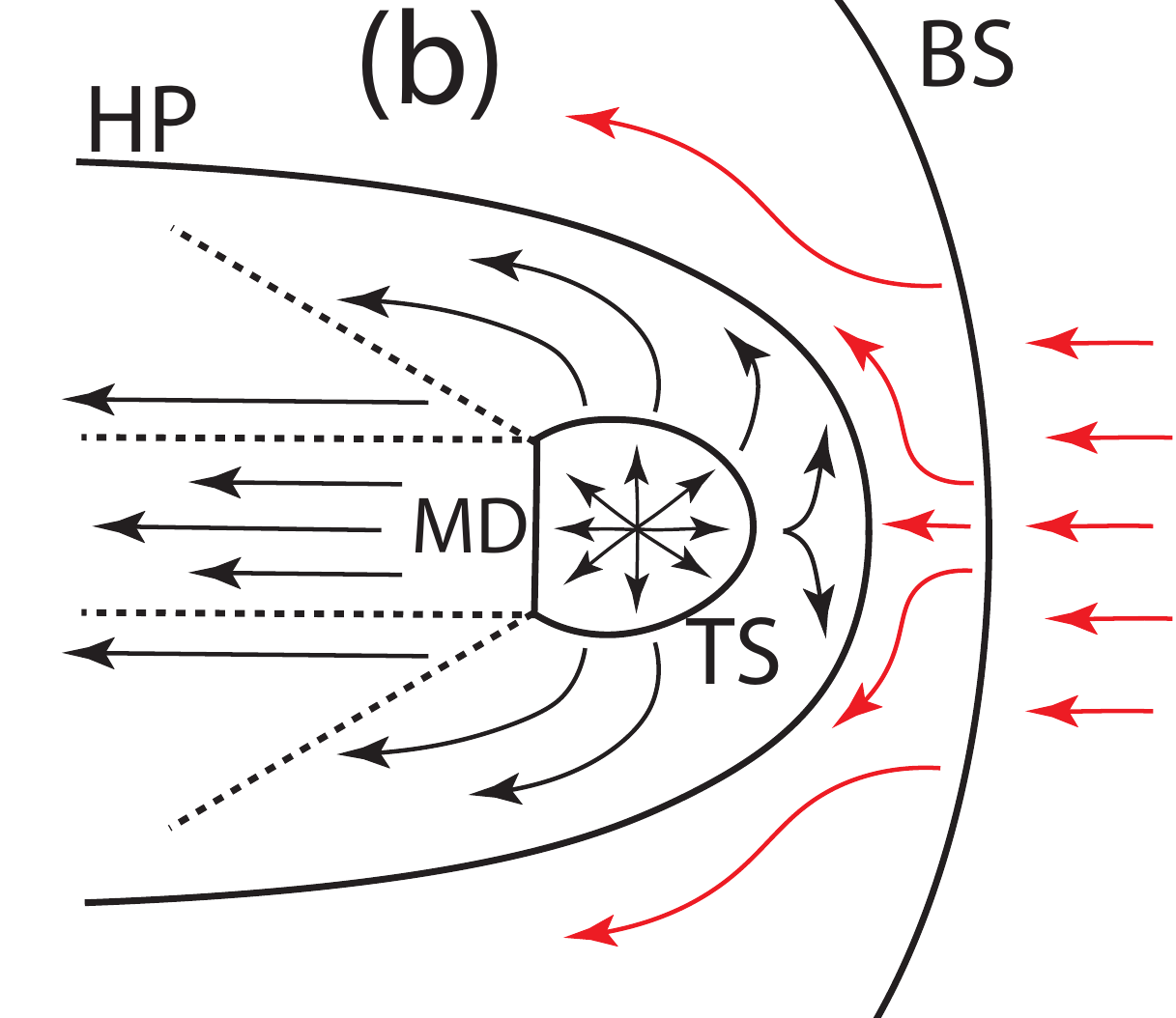}
	\includegraphics[width=0.33\textwidth]{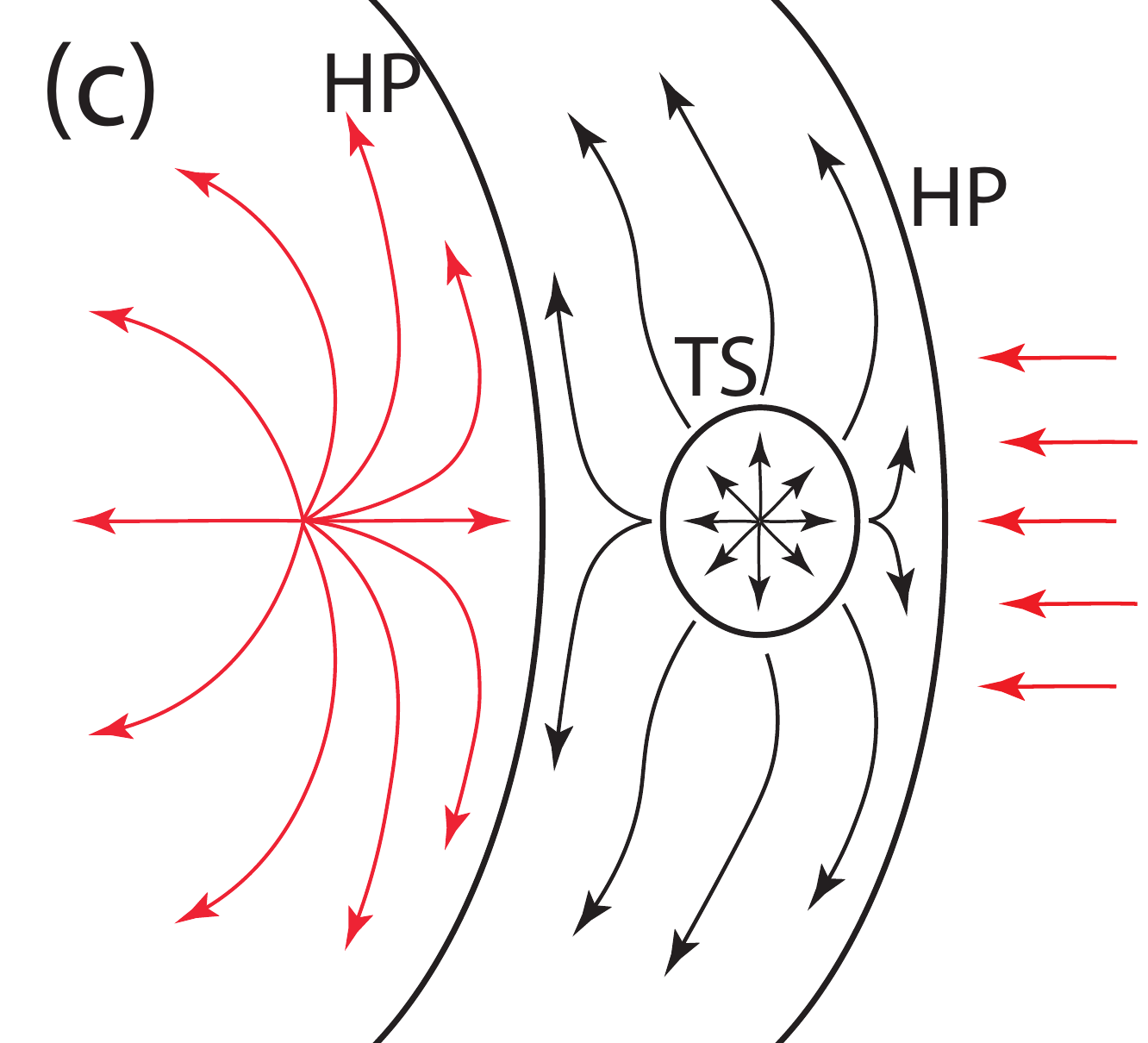}
	\caption{Schematic picture of the heliospheric interface with a sheet-like topology of the heliopause (HP) for small (a) and large (b) Mach numbers of the interstellar flow and a tube-like topology (c).}
	\label{astrosphere_sketches}
\end{figure*}

\cite{parker61}  presented an analytical stationary solution of the solar wind (SW) interaction with the slowly moving interstellar medium (ISM) with respect to the Sun. Mach number of the interstellar flow was assumed to be much smaller than one ($M_{\infty}<<1$) , and, therefore, the flow was assumed to be not compressible and potential. In this solution the shape of the heliopause, which is the tangential discontinuity separating the solar wind from the interstellar gas, was obtained. In particularly, it was shown that the heliopause surface is opened towards the tail direction and far from the Sun it approaches a cylindrical surface with radius 2$L_0$, where $L_0$ is the heliopause stand off distance in the direction towards the interstellar flow that we call the upwind direction hereafter.  \cite{baranov70} have considered another extreme case of the fast (hypersonic, $M_{\infty} >>1$) moving interstellar medium. In this case the structure of the interaction region is determined by three discontinuities: 1) the heliopause, 2) the heliospheric termination shock (TS) at which the supersonic solar wind is decelerated to subsonic, 3) the bow shock (BS) at which supersonic stellar wind is decelerated to subsonic.
Assuming hypersonic flow for the stellar wind the problem has been solved in the Newtonian thin layer approximation when the thickness of the interaction region (between TS and BS) is much smaller than its distance to the Sun. In particularly, it has been shown that the shape of the heliopause
is $\sim \alpha/sin \alpha$, where $\alpha$ is the angle calculated from the upwind direction.
Later the structure of the heliosphere in the tail direction, i.e. in the direction opposite to upwind, has been studied numerically.
In particularly, it has been shown \cite{Baranov_Malama_1993} that a quite complex shock wave structure including Mach disk (MD), reflected shock (RS), secondary tangential discontinuity (TD) (Fig.~\ref{astrosphere_sketches}(b)) is formed.

As it has been known since 1980s the interstellar hydrogen atoms penetrate the heliosphere and strongly interact by charge exchange with both the interstellar and solar wind plasma. Figure 2 in \cite{Izmodenov_2000} clearly demonstrates that the heliocentric distances to the termination shock and the heliopause are reduced by a factor of two or more in the self-consistent two-component model of \cite{Baranov_Malama_1993} as compared with  gasdynamic 'plasma only' model. Despite strong dynamical effect the interaction does not influence the geometrical pattern of the flow in the interaction region.  \cite{alexashov_izmodenov2003} explored how charge exchange influences the plasma flow in the tail region. It has been shown that the jump of the plasma parameters across the heliopause disappears at $\sim$3000 AU counted along the axis determined by the interstellar wind velocity vector.

The other component that strongly influences the SW/ISM interaction region is the interstellar magnetic field (IsMF). The effect of IsMF on the heliopause has been firstly studied by Baranov and Zaitsev (1995) in the axisymmetric case when IsMF is parallel to the interstellar velocity vector, $\mathbf{V}_{ISM}$. Pogorelov and Matsuda (1998) and Linde (1998) have performed 3D MHD calculations of the global heliosphere and have shown the tilted (with respect to the direction of the interstellar gas flow) IsMF leads to asymmetry of the global heliosphere. Izmodenov et al. (2005) have demonstrated in the frame of  the self-consistent 3D kinetic-MHD model that the tilted IsMF makes both the heliopause and the heliospheric termination shock strongly asymmetric (see, also, recent papers by Izmodenov and Alexashov, 2015, 2020). Moreover, the asymmetry in plasma flow around the heliopause causes the deflection of the interstellar H atoms in the heliospheric interface. Such deflection has been discovered in the analyses of SOHO/SWAN H-cell data (see Lallement et al., 2005 Science).   Overall,  both classical and modern 3D multi-component models demonstrate that the heliopause is open in the tail and  it is topologically equivalent to a plane (Fig.~\ref{astrosphere_sketches}(a,b)). 


Recently, this 'classical' view on the global structure of the heliopause underwent a major revision.
\cite{opher15} and \cite{drake15} have shown that the heliopause may have a tube-like (or 'croissant')  shape (Fig.~\ref{astrosphere_sketches}c) due to the influence of the azimuthal component of the heliospheric magnetic field. 
 In fact, the idea that the solar wind flow splits into two streams due to azimuthal magnetic field has been investigated about 50 years ago by \cite{Yu1974}. However, this paper has been forgotten and the same idea was reappeared in 2015. A tube-like shape has been obtained  in \cite{opher15} in the frame of their numerical three-dimensional multi-fluid magneto-hydrodynamic (MHD) code for the case when the interstellar gas flows with respect to the star.  It is important to note that the 'croissant' shape was obtained by \cite{opher15} for the case of unipolar solar magnetic field. This allows to reduce numerical dissipation of the magnetic field  in the models with varied polarity and current sheet.

The jet formation due azimuthal magnetic field was also clearly demonstrated by \cite{drake15}, \cite{golikov2017a}, \cite{golikov2017b} for the case when the solar wind flows into the homogeneous interstellar gas at rest. 

To give a simple qualitative illustration  of the azimuthal magnetic field effect we can estimate the magnetic force $\mathbf{F}_{mag} = ([\nabla \times \mathbf{B}] \times \mathbf{B})/(4\pi)$ in the region beyond the termination shock by using the solution of a classical problem when the flow of supersonic point source interacts with surrounding interstellar at rest. This solution includes the termination shock at $R_{TS} \sim \sqrt{\frac{\dot{M}V_0}{4\pi p_{\infty}}}$, where $R_{TS}$ is the heliocentric distance to the termination shock, $\dot{M}$ is the stellar mass loss rate, $V_0$ is the terminal velocity of the supersonic stellar wind, $p_{\infty}$ is the interstellar gas pressure. In the supersonic SW (for $R<R_{TS}$) the solution is $V \sim V_0$, $\rho \sim 1/R^2$ and $p \sim 1/R^{2\gamma}$, where $R$ is the distance to the Sun/star. In the subsonic region ($R>R_{TS}$) the gas may be considered as not compressible and
the solution is $V \sim 1/R^2 $, $\rho \sim \rho_{\infty} $ and $p \sim p_{\infty}$.

This solution can be used to calculate the frozen-in magnetic field in the kinematic approximation. Solving the following equation $\nabla \times [\mathbf{V} \times \mathbf{B}]=0$  and assuming that magnetic field is parallel to the stellar wind velocity vector at the Sun: 
\begin{equation}
R<R_{TS}: B_R \sim 1/R^2,  
\quad B_\phi \sim (1/R)\sin\theta, \quad B_\theta = 0.
\label{parker_field}
\end{equation}
The solution above for $R<R_{TS}$ has been obtained in \cite{parker58}.
\begin{equation}
R > R_{TS}: \quad B_R \sim 1/R^2, \; B_\phi \sim R\sin\theta, \quad B_\theta = 0.
\label{pmf}
\end{equation}
Here $\theta$ is the polar angle counted from the stellar rotational axis (x-axis), $\phi$ is the azimuthal angle.


As it follows from the formula above, the plasma $\beta$ grows with the distance and, therefore, a strong influence of the magnetic field on the plasma can be expected. 
Magnetic force $\mathbf{F}_{mag}$ has the main component in $r$-direction (in the cylindrical ($z$, $r$, $\phi$) coordinate system where $z$-axis is the axis of stellar rotation): $F_{mag, r} = -2r$ for the magnetic field (\ref{pmf}), so the magnetic field  deflects the stellar wind from the original radial direction and flows towards $z$-axis. Hence the bent tube shape structure of the flow is formed. Note that even a weak magnetic field of a star is sufficient to form a tube in the absence of an incident flow (\cite{golikov2017a, golikov2017b}).

The problem of the magnetized stellar wind flow into the interstellar medium at rest has been explored in details by \cite{golikov2017a, golikov2017b}. It has been shown that in dimensionless form the solution depends on only one parameter that is the Alfvenic Mach number. The parametric study has been performed both analytically and numerically making the results double justified.  According to the numerical results the distances from the axis of stellar rotation  to the heliopause at the equatorial
plane (z = 0) and in the jets (x $\rightarrow \infty$) are both proportional to $M_A^{1/3}$ as long as $M_A$ is large enough ($M_A>10$). 

The bent-tube (or "croissant") shape of the heliopause  has been disputed by \cite{pogorelov2015}. The first paper argues that the bent tube-shape structure in the tail disappears due to solar cycle effects, and, in particularly due to change of the angle between solar magnetic and rotation axes. In addition, it is argued the effects of charge exchange of protons with the interstellar hydrogen atoms at kinetic level as well as Kelvin-Helmholtz instabilities may destroy the bent tube shape. \cite{Pogorelov2017SSR} have shown that in the time-dependent model with the solar cycle the bent-tube disappears even though the heliospheric magnetic field is unipolar.

\cite{izmodenov_alexashov2015} demonstrated in their 3D kinetic-MHD model that the heliopause has a shape that is topologically equivalent to a plane as in the classical models (in opposite, bent-tube shape is topologically equivalent to cylinder).  Although, of course, the heliopause shape is distorted by the unipolar heliospheric magnetic field as it should. The paper also demonstrates that dynamic effect of the heliospheric magnetic field exists and it is pronounced in the increase of the solar wind flux at higher latitudes in the tail. The differences of these results with \cite{opher15} were associated with different treatments of the interstellar neutral component. Another discussed possibility is that the plane-topology obtained in \cite{izmodenov_alexashov2015}
could be induced by the specific numerical grid. In this paper we will show that all these arguments are not necessary needed to have plane-like heliopause.  

In this work we continue the study started in \cite{golikov2017b} on exploration how the azimuthal stellar magnetic field effects 
on the global structure of a simplified ('toy') ideal-MHD astrosphere.
However, here we consider the more general case when interstellar gas moves with respect to the star.


The structure of the paper is the following. Section 2 presents a description of the model and formulates the assumptions that we extended. The basic MHD equations are set and the used boundary conditions are written out. In subsection 2.1 we formulate the problem in dimensionless form.  In Section 3  numerical approach is discussed. The spatial grid and methods for solving the problem of discontinuity decay are described. Section 4 presents and discusses the results of a numerical study of the problem. Finally, Section 5 summarizes this work, and discusses issues for the future research.

\section{Model}

The goal of this paper is to explore the effects of the azimuthal stellar magnetic field on the global structure of the stellar/interstellar wind interaction in the frame of the most simple approach. 
We restrict ourselves by the following assumptions:
\begin{itemize}
    \item both the stellar wind and ISM are considered in the frame of one-fluid ideal MHD approach. In other words, we neglect all possible physical dissipating processes. We assume that both gases are fully ionized hydrogen plasma, so the pressure and the temperature are linked as $p=2 n_p k_B T$, where $n_p$ is the proton number density,  $k_B$ is the Boltzmann constant, $T$ is the plasma temperature. The ratio of specific heat fluxes, $\gamma$,  is assumed to be equal to 5/3;
    \item the interstellar magnetic field is assumed to be zero;
    \item the stellar wind at the inner boundary of the computational domain is assumed to be spherically-symmetric and supersonic;
    \item the azimuthal component of the stellar magnetic field  at the inner boundary is assumed to be Parker spiral solution:
\begin{equation}
\label{spiral}
~B_\varphi= B_{\varphi,E} \left(\frac{R_E}{R}\right)sin\theta. 
\end{equation} 
\end{itemize}

Radial component $B_R$ is assumed to be zero. This assumption is not critical since $B_R\sim1/R^2$ is small as compared to $B_\varphi$ at large heliocentric distances. We pose the conditions to keep symmetry of the solution with respect to the $y=0$ and $z=0$ planes of the Cartesian coordinate system with x-axis directed toward the interstellar wind flow, z-axis directed toward the axis of stellar rotation, y axis oriented to complete right-handed coordinate system. $(R,\theta,\varphi)$ are the spherical coordinates connected with the 
solar equatorial plane: $\theta$ is the solar latitude, counted from
north solar pole ($0^{\circ}$) to south ($180^{\circ}$).

We are interested in steady-state solutions of the SW/ISM interaction region. The governing equations are
\begin{equation}
\nabla \cdot( \rho \mathbf{V})=0,
\label{eq-continuity-MHD-section}
\end{equation}
\begin{equation}
\label{eq-momentum-MHD-section}
\nabla\cdot\left[\rho{\mathbf{V}}{\mathbf{V}} + \left(p + \frac{B^2}
{8\pi}\right){\mathbf{I}} - \frac{{\mathbf{B}}{\mathbf{B}}}
{4\pi}\right] = 0,
\end{equation}
\begin{equation}
\label{eq-MF-MHD-section}
\nabla\cdot({\mathbf V}{\mathbf B} - {\mathbf B}
{\mathbf V})=0,
~~\nabla \cdot \mathbf{B} = 0,
\end{equation}
\begin{equation}
\label{eq-energy-MHD-section}
\nabla\cdot\left[\left(E + p + \frac{B^2}{8\pi}\right){\mathbf V} -
\frac{({\mathbf V} \cdot {\mathbf B})}{4\pi}{\mathbf B}\right]
= 0,
\end{equation}
 where  ${\mathbf B}$ is the magnetic field induction vector,
	${\mathbf a}{\mathbf b}$ is the tensor production of two vectors ${\mathbf a}$ and
	${\mathbf b}$, ${\mathbf I}$ is the unity tensor, '$\cdot$' is the scalar product,
	$E=\frac{\rho V^2}{2} + \frac{p}{\gamma - 1} + \frac{B^2}{8\pi}$.

To finish the formulation of the problem, the inner boundary conditions in the unperturbed (by ISM) SW and the outer boundary conditions in the pristine ISM should be formulated.
At the outer boundary in the unperturbed ISM the density $\rho_{\infty}$, velocity $\mathbf{V}_{\infty}$, pressure $p_{\infty}$  are assumed to be known.
For the inner boundary conditions, we use the hyper-sonic solution for the stellar wind 
that is determined by the stellar mass loss rate $\dot{M}_s$, the terminal velocity of the stellar wind, $V_0$, and $F_B = B_{\varphi,E} R_E$ is the constant determined by the stellar magnetic field, where $B_{\varphi,E}$ is the strength of the magnetic field at the distance of $R_E$. 

\subsection{Dimensionless formulation and parameters}

The solution of the problem formulated above depends on the following seven parameters:
\begin{equation}
V_0, \  K = \rho_E R^2_E V^2_0,  \  F_{B} = B_{\varphi,E} R_E,  \  \rho_{\infty}, \ p_{\infty},  \  V_{\infty}, \  \gamma.
\label{parameters}
\end{equation}
Here $V_{\infty}$ is the magnitude of the ISM velocity; $\rho_E$ and $B_{\varphi,E}$ are the density and magnetic field of the stellar wind at the distance $R_E$, which can be chosen arbitrary with the restriction that the stellar wind speed reaches its terminal value of $V_0$ at that distance. The parameters $K$ and $F_B$ do not depend on the choice of $R_E$ because $K= V_0  \dot{M}_\odot/(4 \pi)$ and $B_\varphi R = const$ in the Parker spiral.

To reformulate the problem in the dimensionless form we choose the following three characteristic parameters:
1) $\rho_{\infty}$ as the characteristic density, 2) the ISM sound speed, $a_{\infty}=\sqrt{\gamma p_{\infty}/\rho_{\infty}}$, as the characteristic velocity, and 3) the characteristic distance $R_* = \sqrt{\frac{K}{\gamma p_\infty}}$. This distance is proportional to the termination shock stand-off distance obtained from the analytical stationary solution of the problem of the non-magnetized spherically symmetric stellar wind interaction with ISM at rest (\cite{parker61}).  We did not choose $V_{\infty}$ as a characteristic speed because limiting case $V_{\infty} =0$ will be considered as well.

In dimensionless formulation the seven parameters (\ref{parameters}) turn into the following dimensionless parameters:
\begin{align}
 \frac{V_0}{a_\infty} ,\  1,\ \frac{\sqrt{4 \pi} }{M_A},\ 
 1,\  \frac{1}{\gamma},\  M_\infty,\ \gamma,
\end{align}
respectively.
$M_A = V_0\sqrt{4 \pi \rho_E}/B_E$ is Alfvén Mach number in the stellar wind, $M_\infty$ is gasdynamic Mach number. Further in the paper the ratio $\frac{V_0}{a_\infty}$ is denoted as $\chi$.

The parameter $\chi$ is non essential for the stationary problem considered here. Indeed, let $\rho_1(\mathbf{r})$, $\mathbf{V}_1(\mathbf{r})$, $p_1(\mathbf{r})$, $\mathbf{B}_1(\mathbf{r})$  be the solution of (\ref{eq-continuity-MHD-section})-(\ref{eq-energy-MHD-section}) 
for some value of $\chi=\chi_1$. Then for $ \chi= \chi_2$ (and for the other dimensionless parameters remaining to be the same) let us construct the functions 
$\rho_2(\mathbf{r})$, $\mathbf{V}_2(\mathbf{r})$, $p_2(\mathbf{r})$, $\mathbf{B}_2(\mathbf{r})$ in such way that out of the tangential discontinuity (i.e. in ISM) they coincide with $\rho_1(\mathbf{r})$, $\mathbf{V}_1(\mathbf{r})$, $p_1(\mathbf{r})$, i.e. the solution remains the same as for  $\chi_1$.
In the stellar wind we assume
\begin{eqnarray}
\rho_2(\mathbf{r}) = \frac{\chi^2_1}{\chi^2_2} \rho_1(\mathbf{r}),\ 
\mathbf{V}_2(\mathbf{r}) = \frac{\chi_2}{\chi_1} \mathbf{V}_1(\mathbf{r}), \nonumber \\
 p_2(\mathbf{r}) = p_1(\mathbf{r}),\ \mathbf{B}_2(\mathbf{r})=\mathbf{B}_1(\mathbf{r}).
\end{eqnarray}

Functions  $\rho_2(\mathbf{r})$, $\mathbf{V}_2(\mathbf{r})$, $p_2(\mathbf{r})$, $\mathbf{B}_2(\mathbf{r})$ satisfy: 1) differential equations (\ref{eq-continuity-MHD-section})-(\ref{eq-energy-MHD-section}) ;  2) Rankine-Hugoniot conditions at the shocks;
3) pressure balance and $(\mathbf{V} \cdot \mathbf{n}) = 0$, $(\mathbf{B} \cdot \mathbf{n}) = 0$ conditions at the tangential discontinuity, $\mathbf{n}$ is the normal to the tangential discontinuity surface; 4) the inner boundary condition $V_0 = \chi_2  a_{\infty}$.
By this we have shown that the geometric pattern of the flow does not depend on $\chi$, and the solution can be obtained for different $\chi$ by a simple re-normalization. Therefore, the parameter  $\chi$ is not essential for the stationary solutions.

Finally, in dimensionless form the problem depends only on the three parameters: 1) gasdynamic Mach number in the ISM ($M_{\infty}$), 2) Alfvenic Mach number in the stellar wind ($M_A$) and 3) the parameter $\gamma$. In section 4 we present the solution of the considered problem for different values of the dimensionless parameters $M_{\infty}$ and $M_A$. The parameter $\gamma$ is kept the same and equal to 5/3 for fully ionized plasma.

It is important to note, that in
the numerical calculations, we have to put inner boundary conditions at a certain dimensionless distance $\hat{R}_{in}$ from the star. In the performed calculations we employed $\hat{R}_{in} = 0.07$.
In the dimensionless form the  boundary conditions at $\hat{R}_{in}$ are the following

\begin{eqnarray}
\label{boundary_conditions_1ae}
V_{in} = V_0/a_{\infty} = \chi, 
\rho_{in} = \frac{K}{ V^2_0 {R}^2_{in} R^2_* \rho_{\infty}} =
\frac{1}{\chi^2 \hat{R}^2_{in}},   
    B_{in} = \frac{\sqrt{4 \pi}}{\hat{R}_{in} M_A }, \nonumber 
\end{eqnarray}


To estimate the dimensionless parameters for the Sun we assume the following values at $R_E$= 1 A.U.: $V_E$ =  432  km s$^{-1}$,  n$_{p, E}$ = 6 cm$^{-3}$, $B_E$ = 37.5 $\mu$ G, V$_\infty$ = 26.4 km s$^{-1}$, T$_\infty$ = 6530 K, n$_{\infty}$ = 0.04 cm$^{-3}$. 
For dimensionless parameters we have $M_\infty = 1.968$, $M_A = 12.937$, $\chi = 32.2$,
The characteristic distance is $R_* = 394.5$ AU.

\section{Numerical approach}

The solution of the stationary equations (\ref{eq-continuity-MHD-section})-(\ref{eq-energy-MHD-section}) is obtained by solving corresponding non-stationary MHD equations with stationary boundary conditions formulated above. The solution of the stationary problem is obtained at $t \rightarrow \infty$. 

Non-stationary MHD equations are solved in the following conservative form in Cartesian coordinates:
\begin{eqnarray}
\frac{\partial\bf U}{\partial t} + \frac{\partial\bf E}{\partial x} + \frac{\partial\bf F}{\partial y} + \frac{\partial\bf G}{\partial z} = 0 \label{6},
\end{eqnarray}

where the following notation for column vectors is introduced: \\
{$\bf U$} = $[\rho, \rho u, \rho v, \rho w, e, B_x, B_y, B_z]^T$,\\
{$\bf E$} = $[ 
\rho u,\
\rho u^2 + p_* - \frac{B_x^2}{4 \pi},\
\rho uv - \frac{B_x B_y}{4 \pi},\
\rho uw - \frac{B_x B_z}{4 \pi},\
(e + p_*)u - \frac{B_x}{4 \pi} {\bf (v\cdot B)},\
0,\
uB_y - vB_x,\
uB_z - wB_x
]^T$,\\
{$\bf F$} = 
$[\rho v,\
\rho uv - \frac{B_x B_y}{4 \pi},\
\rho v^2 + p_* - \frac{B_y^2}{4 \pi},\
\rho vw - \frac{B_y B_z}{4 \pi},\
(e + p_*)v - \frac{B_y}{4 \pi} {\bf (v\cdot B)},\
vB_x - uB_y,\
0,\
vB_z - wB_y]^T$,\\
{$\bf G$} = $[ \rho w,\
\rho uw - \frac{B_x B_z}{4 \pi},\
\rho vw - \frac{B_y B_z}{4 \pi},\
\rho w^2 + p_* - \frac{B_z^2}{4 \pi},\
(e + p_*)w - \frac{B_z}{4 \pi} {\bf (v\cdot B)},\
wB_x - uB_z,\
wB_y - vB_z,\
0]^T$,\\
 where $p_* = p + \frac{{\bf B}^2}{8\pi}$ is the total pressure.\\

Note that nonzero divergence of the magnetic field may appear in numerical calculations. This happens for two reasons: (1) inaccurate numerical scheme, especially at the discontinuities, (2)  incorrect initial conditions for the magnetic field. \cite{Powell_1999} proposed a method of divergence cleaning. 
According the method,  an additional vector $\bf P$  proportional to div${\bf B}$ has been added to the right part of the governing system of equations. So the system (\ref{6}) becomes: 
\begin{eqnarray}
\frac{\partial\bf U}{\partial t} + \frac{\partial\bf E}{\partial x} + \frac{\partial\bf F}{\partial y} + \frac{\partial\bf G}{\partial z} = -{\bf P}, \label{7}
\end{eqnarray}
where ${\bf P} = \text{div}{\bf B} \cdot [0,\ \frac{B_x}{4 \pi},\ \frac{B_y}{4 \pi},\ \frac{B_z}{4 \pi},\ \frac{(\bf v\cdot B)}{4 \pi},\ u,\ v,\ w]^T$.\\
Applying divergence to the Faraday equation (i.e. to the last three equations of (\ref{7}) written in vector-form), we obtain:
$$\frac{\partial (\text{div}{\bf B})}{\partial t} + \text{div}({\bf v} \text{div}{\bf B}) = 0.$$
This is the transport equation for the divergence. It demonstrates that initially non-zero divergence of the magnetic field is convected out of the computational domain with time.

Together with the system (\ref{7}) we solve the linear 
transport equation:
$$ \frac{\partial \rho Q}{\partial t} + \text{div}({\bf v} \rho Q) = 0 ,$$
where $Q$ is an indicator that is passively transported together with the gas (e.g. \cite{Osher}).
If we set, for example, Q  = 1 for the stellar wind and Q (r) = 100 for the ISM material, then the solution of the transport equation allows us to determine the shape of the heliopause quite precisely.

\begin{figure} 
 \begin{subfigure}{0.205\textwidth}
		\centering
		\includegraphics[width=1\linewidth]{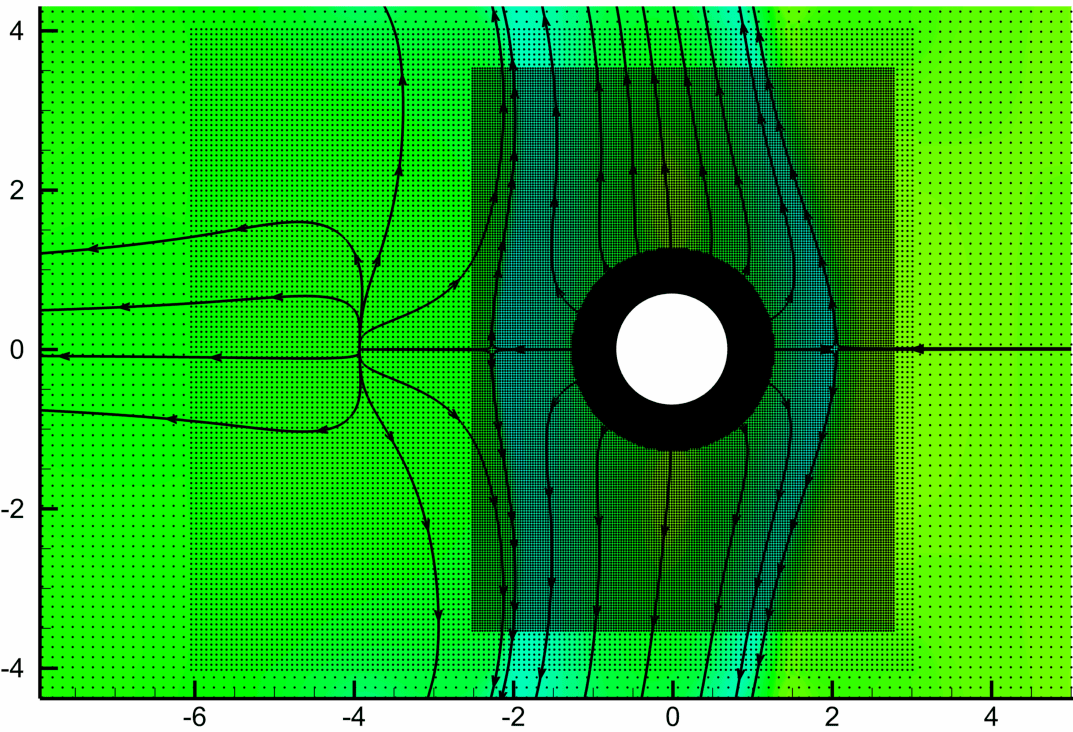}  
		\label{fig:sub-first1}
	\end{subfigure}
	\begin{subfigure}{0.23\textwidth}
		\centering
		\includegraphics[width=1\linewidth]{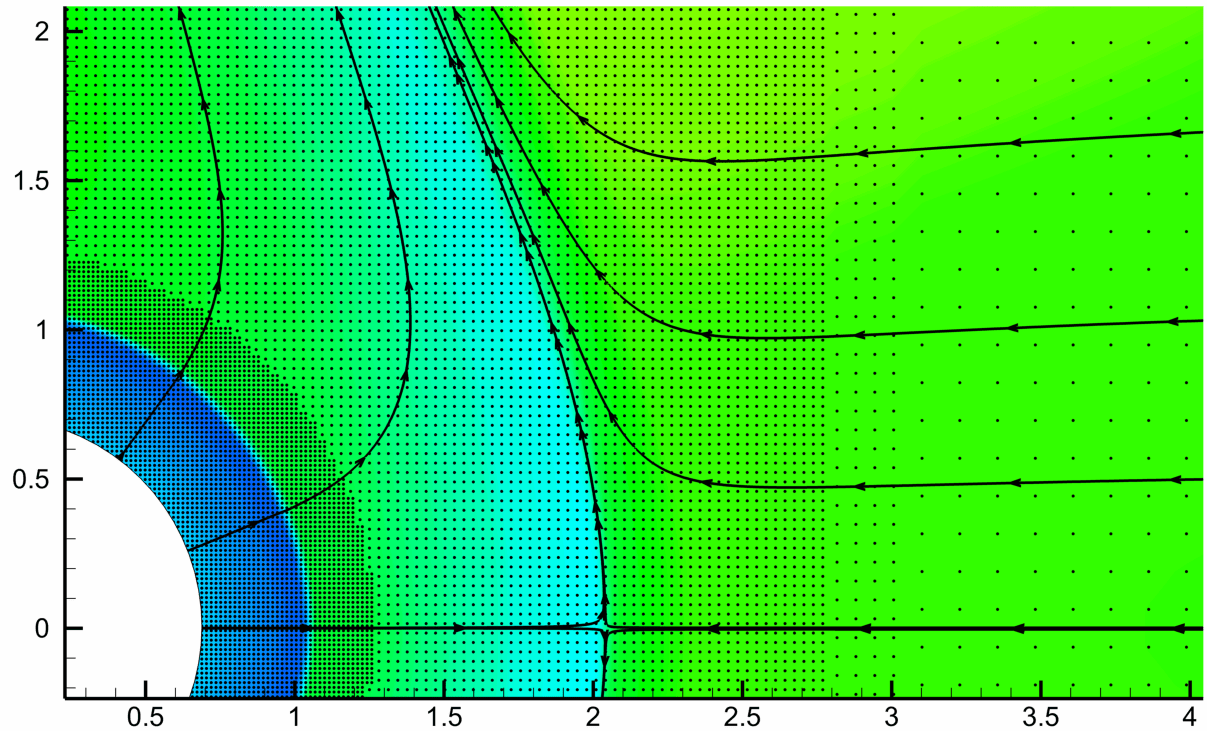}  
		\label{fig:sub-first5}
	\end{subfigure}
	\caption{The example of mesh refinement employed for the results of this paper. The dots mark the centers of the Cartesian cells.}
	\label{exp}
\end{figure}

 We divide the computational domain into a regular Cartesian grid consisting of rectangles.
The grid employed for this paper has about 10 million cells.
Our numerical code allows to perform mesh refinement in  sub-domains specified by user. We performed four levels of mesh refinement. At each level, every cell of the chosen sub-domains is divided by eight smaller cells. Thus, after one level of mesh refinement the grid resolution becomes twice as good in all spatial directions. Figure \ref{exp} shows an example of a grid used by us in the calculations presented in this paper. The dots present the centers of rectangular cells. (The plasma streamlines and density are shown in this figure just to demonstrate locations of the discontinuities.) Note that some numerical and visualization artefacts can appear at the junctions of the grid sub-domains with difference resolution.
To minimize the numerical effects, we follow the rules: 1) the junctions of the numerical sub-domains should not be in the flow region with large gradients of gas-dynamic parameters, 2) the spatial resolution of the two sub-domains at junctions should not be different by more than twice (in each direction).


To solve the system of equations (\ref{7}) numerically  we used the HLLC-type Riemann solver (e.g. \cite{Gurski}). The solver was constructed under the assumption that the normal component of velocity is constant over the Riemann rarefaction wave. The HLLC-type Riemann solver cannot simulate the structures related to the Alfve´n and slow waves with high enough resolution although isolated contact discontinuities as well as isolated fast shocks can be resolved exactly.

The first order of approximation in space and time was used for the results presented in this paper, i.e. the plasma parameters were assumed to be constant within each of computation cells. The “second” order scheme with flux limiters (e.g. minmod) has been employed as well. The results of the first and second order schemes are quite similar, in fact. However, in their results we see some artificial noise in isolines of the plasma parameters at the borders of domains with different spatial resolution. To avoid possible confusion of the readers we present only the first order scheme results.

At the outlet boundaries we use so-called soft boundary conditions when derivatives of all values are assumed to be zero.  It is quite known that such conditions can lead to the formation of reverse flow and vortex in the tail region. To avoid these problems we introduce some small outlet flow at the boundary. This condition works at the initial stages of time relaxation, while at the final stages it is replaced by the soft boundary conditions, so, final stationary solution satisfies the soft-boundary conditions as it is declared above.

Finally,  at all inlet boundaries, we fix the values according to the boundary conditions. Mathematically this approach is not strictly correct because in this case one of the gasdynamic characteristics arrives at the boundary from inside of the computational domain. To avoid the influence of the outcoming characteristics we performed specific numerical calculations by increasing the computational domain in the upwind direction significantly. For such a large domain the perturbations coming from inside will dissipate (due to numerical scheme dissipation) and do not influence the parameters at the outer boundaries. Such calculations give us additional validation of our numerical results for low Mach numbers.

The calculations were carried out using a GPU device: GeForce GTX 1080 Ti and Nvidia CUDA compiler (nvcc). This allows to speed up calculations in about five times as compared to programming with 16 CPU threads (\cite{Korolkov_2020}).

\section{Results}

We performed parametric study by varying the two dimensionless parameters of the problem, $M_A$ and $M_{\infty}$.
The results are presented in the Cartesian coordinate system introduced above: the $x$-axis is directed toward the interstellar gas flow, $z$-axis is coincided with the axis of stellar rotation. $y$-axis completes the right-handed coordinate system.

\begin{figure*}
	\begin{subfigure}{1.0\textwidth}
		\centering
		\includegraphics[width=1.0\linewidth]{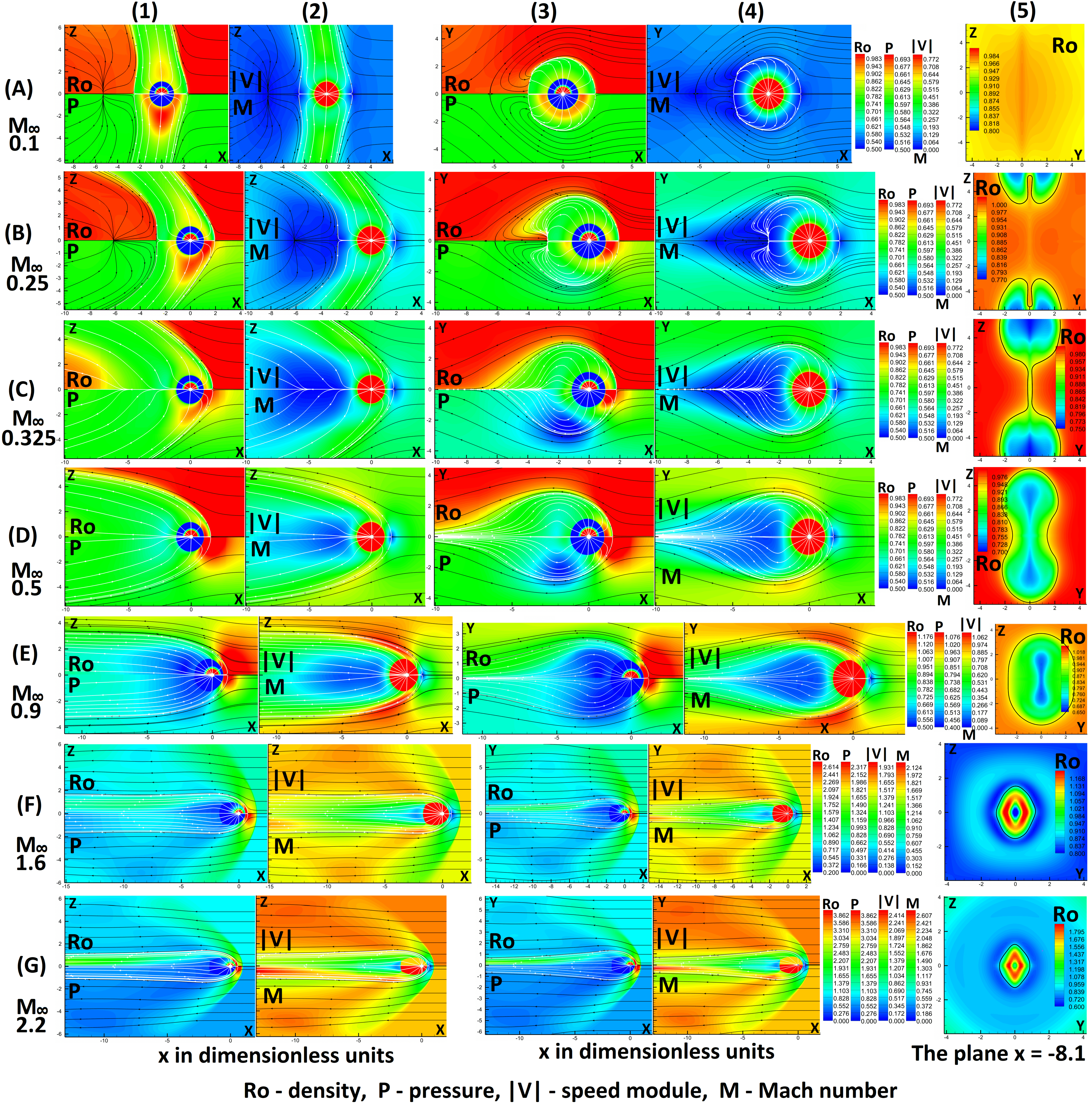}  
		\label{fig:sub-first3}
	\end{subfigure}
	\caption{  Distribution of flow parameters in three planes: $y = 0,\ z = 0,\ x = -8.1$ \ for $\ M_A = 12,\ \chi = 2$,\ HLLC-type\ method.}
	\label{fig_M12}
\end{figure*}

\begin{figure*}
	\begin{subfigure}{1.0\textwidth}
		\centering
		\includegraphics[width=0.93\linewidth]{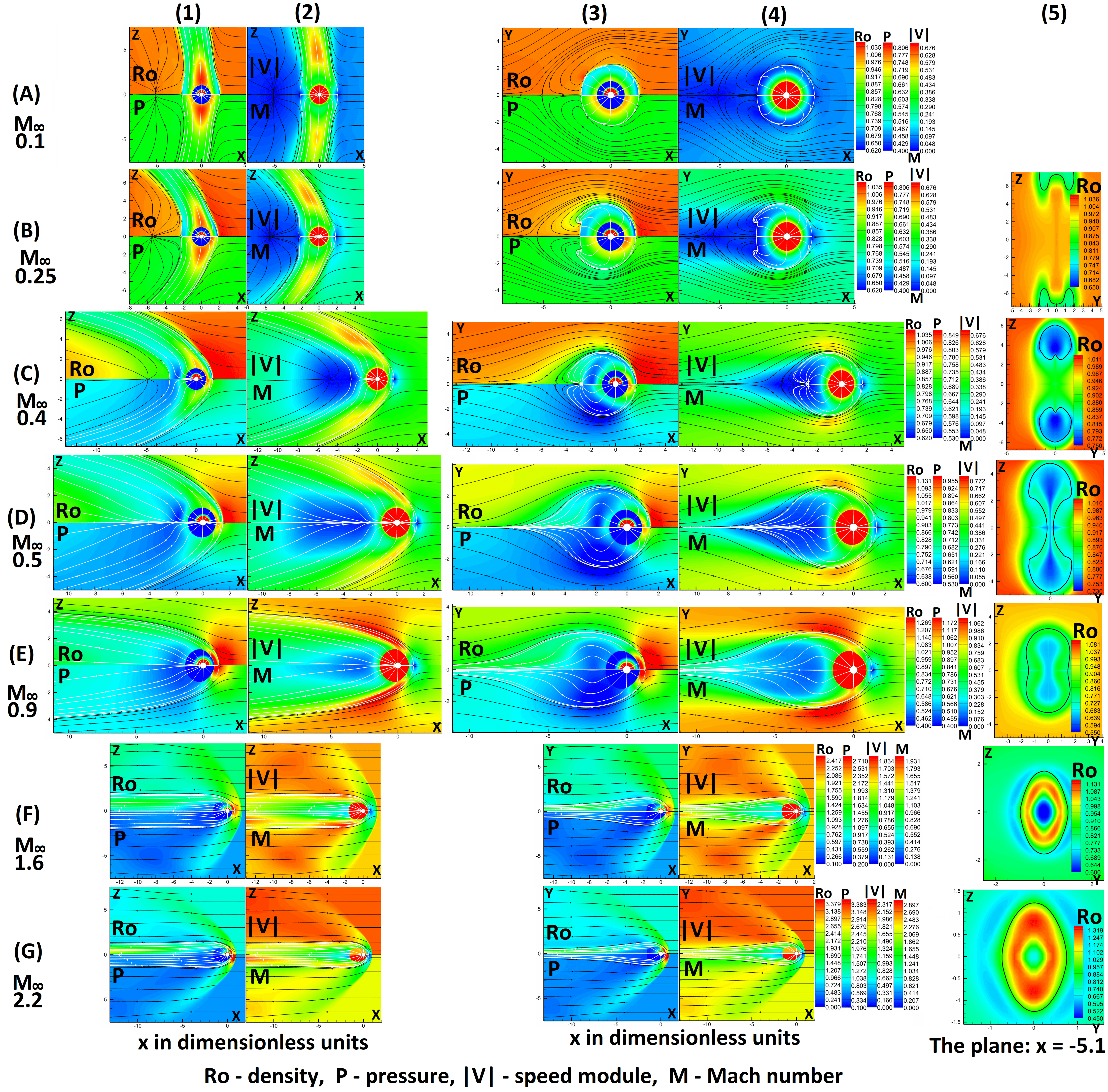}  
		\label{fig:sub-first4}
	\end{subfigure}
	\caption{  Distribution of flow parameters in three planes: $y = 0,\ z = 0,\ x = -5.1$ \ for $\ M_A = 8,\ \chi = 2$,\ HLLC-type\ method.}
	\label{fig_M8}
\end{figure*}

\begin{figure*}
	\begin{subfigure}{1.0\textwidth}
		\centering
		\includegraphics[width=0.93\linewidth]{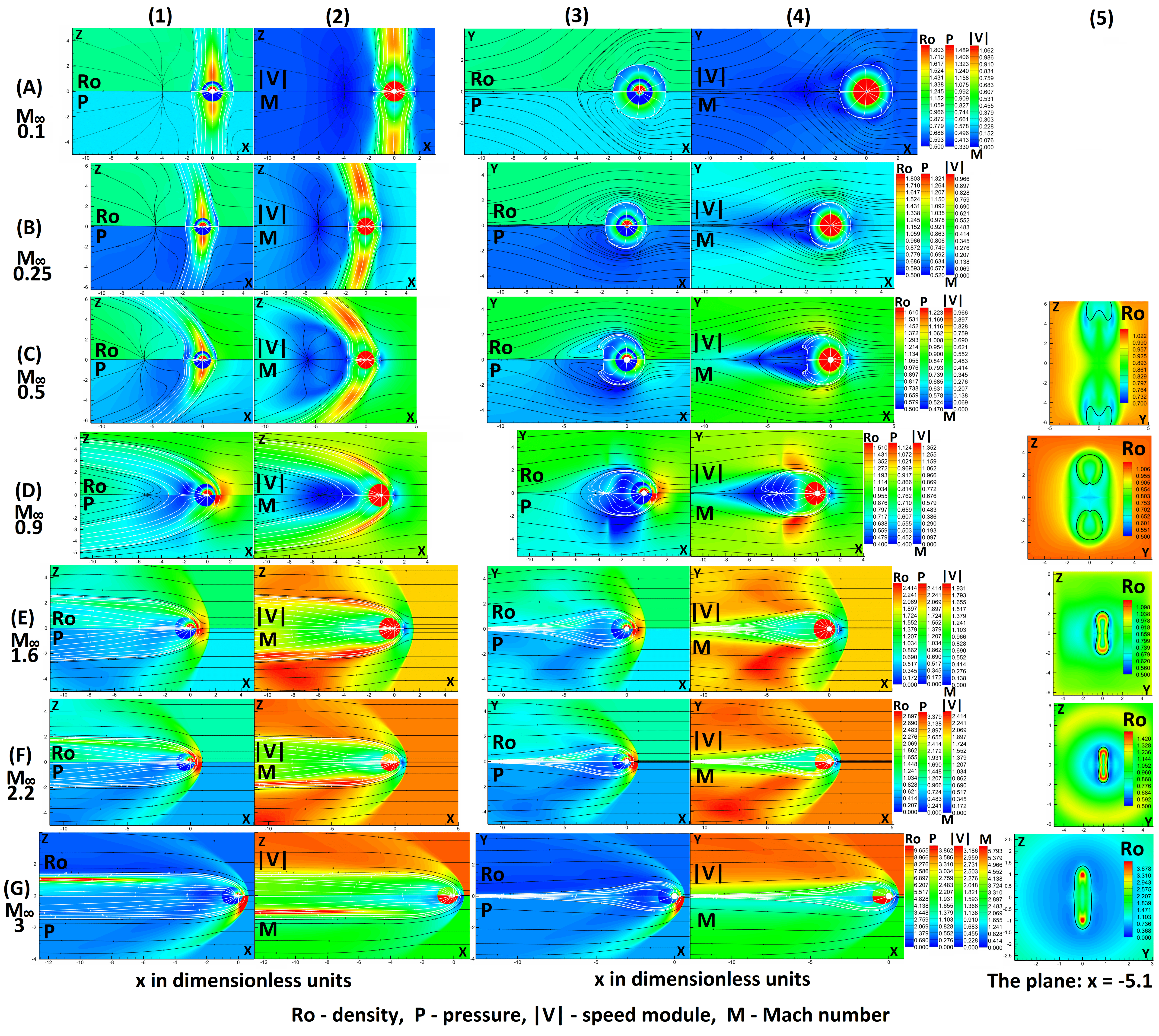}  
		\label{fig:sub-first2}
	\end{subfigure}
	\caption{  Distribution of flow parameters in three planes: $y = 0,\ z = 0,\ x = -5.1$ \ for $\ M_A = 4,\ \chi = 2$,\ HLLC-type\ method.}
	\label{fig_M4}
\end{figure*}

Figures  \ref{fig_M12}, \ref{fig_M8}, \ref{fig_M4} present the results of the calculations for $M_A =12, 8, 4$, respectively. 
Each of the figures has five columns and seven rows. Each row (marked by letter from A to G) presents results of calculation with a certain value of Mach number of the interstellar flow, $M_{\infty}$. The value of $M_{\infty}$ varies from 0.1 for the results in row (A) to 2.2 in row (G). The latter value is close to the heliospheric value of 2.17. 

For each set of model parameters, the results are presented in three panels: 1)  $y=0$ (panels in columns (1) and (2) in Figures) that is the meridian plane containing the stellar rotation axis and ISM inflow axis, 2) $z=0$ (columns (3) and (4) in Figures) that is the equatorial plane, and 3) $ x = const$ (columns (5)) that is the plane perpendicular to the direction of the interstellar flow. The value of the constant is different at different plots.  Note, that the first two planes are the planes of symmetry due to simplified assumptions made in our model.

The numerical results are presented as follows: 1) the plasma density is shown in the upper parts of each panel in columns (1) and (3), 2) the total pressure, i.e. the sum of thermal plasma pressure and the magnetic field pressure, is shown in the bottom parts of each panel in columns (1) and (3), 3) the absolute value of plasma velocity is shown in the upper parts of each panel in columns (2) and (4),
4) the Mach number is shown in the bottom parts of each panel in columns (2) and (4), 5)  column (5) shows the density distribution in the plane perpendicular to the ISM flow vector.  Also, the tangential discontinuity of astropause is shown as solid black line. We determine the discontinuity in such a way that the value of the introduced in the previous section marker $Q$ is equal to 30. Note, that the grid resolution in the tail part of the flow is less than it is in the upwind part, so the contact surface is smeared more in the tail. Therefore, the determination of astropause as $Q=30$ is not precise but it is still quite accurate. Each panel presents also the streamlines. The streamlines of the ISM flow are shown as black curves. The  solar wind streamlines are shown as white curves. This representation allows to determine the shape of the astropause for panels in columns (1)-(4).

Below we give our interpretation of the presented numerical results. First of all, our numerical code reproduces the results of \cite{golikov2017a, golikov2017b} paper for $M_\infty =0$. In this case the astropause has a tube-like shape symmetric with respect to z-axis (the axis of stellar rotation).  The results of  $M_\infty =0$  may be found in Figures 6-9 of \cite{golikov2017a, golikov2017b} paper and we do not repeat them here.

Row (A) in  Figure \ref{fig_M12} presents the results in the case of small velocity of the interstellar gas with respect to the Sun. The Mach number is $M_\infty = 0.1$. The global structure of the astrosphere has been changed slightly as compared with the case without flow. 
The bent tube shape remains mostly the same, although it is slightly distorted by the action of the interstellar flow. It is seen in panels (A1) and (A2) in  Figure \ref{fig_M12}. 
The wings of the tube have some tendency to distortion toward the tail. 
The shape of the tube changes not only at the wings but also at the central part of the tube.  This is seen in panels (A3) and (A4) for equatorial ($z=0$) plane. Due to relative motion the ISM pressure at the upwind stagnation point of the astropause is larger compared with the case of the ISM at rest. As a result, the distance to the astropause in upwind is less. Then, the distance to astropause  increases as we move from the upwind stagnation point toward the flanks (in (xy)-plane). This is connected with the pressure drop along the tube. The deformation of the heliospheric tube resembles the deformation of a falling droplet at its initial stage. 
Therefore, the astrospheric tube becomes more elongated toward the y-axis. The elongated tube forms a blunt obstacle for the interstellar flow that creates some rarefaction in the tail region around the downwind stagnation point. For strongly subsonic flows the gas immediately moves toward the rarefaction area and fills the gap.  This is the physical nature of the reverse flow jet that is formed in the tail (Figure  \ref{fig_M12}-A). This nature is quite similar to the formation of the vortexes in the flows around blunt bodies. 
This jet interacts with the astropause at the tail stagnation point and evacuates  along the heliospheric tube toward the north and south directions. Therefore, the vortex does not appear.

Since the general direction of the interstellar flow is toward the tail, a new stagnation point appears in the ISM. It is located 
at about $x=-5$ in the equatorial plane. Since the velocity in the jet is rather small (Figure  \ref{fig_M12}-A2, A4) then all region between the tail stagnation points can be considered as stagnation area.

Row B (panels B1 - B5) of Figure \ref{fig_M12} presents the results of calculations with slightly larger ISM Mach number of $M_\infty$=0.25.
This corresponds to the lager interstellar velocity or smaller  temperature as compared with the case of $M_\infty$=0.1. 
As it is seen from the Figure \ref{fig_M12}, the interstellar wind distorts the astrospheric tube stronger.  The wings of the astrospheric tube have larger deformation towards the tail.  The upwind stagnation points are closer to the Sun. The downwind stagnation point at the astropause does not change its location, while the secondary stagnation point slightly moves out. 
It is also interesting to observe that the streamlines passing through the secondary stagnation point are originated in the interstellar medium closer to the y-axis. Therefore, the reverse flow jet looks more compact. Despite all these differences, the global structure of the flow is qualitatively similar to the flow of $M_\infty = 0.1$. Therefore, we conclude that for small ISM Mach numbers ($M_\infty < 0.3$)  the structure of the magnetised stellar wind interaction with the ISM can be characterised as follows: 1) the bent tube shape of the astrosphere remains, 2) the wings of the tube are distorted toward the tail direction, 3)  the reverse interstellar flow jet is formed in the tail region. The jet flows to the downwind stagnation point at the astropause  and then flows out along the astrospheric tube in the north and the south directions; 4) the secondary stagnation point appears in the ISM in the tail. As we know such a structure of the flows in the SW/ISM interaction has not been discussed so far in literature. 

Rows (C) and (D) in  Figure \ref{fig_M12} present the results obtained for  $M_\infty = 0.325$ and $M_\infty = 0.5$, respectively. It is seen clearly (especially for $M_\infty = 0.5$)  that the structure of the flow in the tail is completely different from the one has been obtained for small Mach numbers. Panels in the first and second columns demonstrate that the astropause is open in downwind. The astrosphere looks quite similar to the "standard" (not tube-like) shape.
Nevertheless, the impression of "standard" shape is still quite illusive. As it is seen in (xy)-plane (panels of third and fourth columns in row D)  the astrosphere in the tail in direction of $y$-axis is narrow compared to its size in $z$ direction. The further we move from the star the narrower the astrosphere (in $y$-axis) gets.  Plot in column (5) of row (C) demonstrates that the astropause has extremely narrow cross connection between two sides of the tube. 
Therefore, the "imprint" of the stellar magnetic field on the global shape of the astropause is still strong despite it does not have a tube-like topology anymore.

To understand how the flow changes from the tube-like astrosphere to the open shape we consider the case of $M_\infty$=0.325 (row C in Figure \ref{fig_M12}) more carefully. The value of 0.325 is close to the critical Mach number at which the flow changes from 'tube-like' to open astropause, i.e. bifurcation of the flow pattern appears. By increasing the ISM Mach number we increase the speed of the two streams that flow around the astropause (as around an obstacle) from two sides, and interact at the secondary stagnation point in the tail. The larger the speed of the interacting streams is the weaker the reverse flow from the stagnation point is, and the smaller the pressure left to the secondary stagnation point is.
At certain value of the ISM Mach number the interstellar pressure can not balance the pressure of the stellar wind. As a result the wind moves out, the primary stagnation point approaches the secondary one, and all together they escape outward.  Therefore, the stellar wind pushes the interstellar gas out to the tail direction. However, it happens in the narrow area around the x-axis. 

The value of $M_\infty$=0.325 is close to the critical value of the flow regime bifurcation between the tube mode and the open heliopause mode. We noticed that the transition itself is happening quite quickly when the parameter $M_\infty$  was changed from 0.30 to 0.325. During the computations with $M_A = 12$ we were unable to catch the exact value of $M_\infty$ of bifurcation. However, for the calculations with $M_A = 4$ presented in Figure \ref{fig_M4} we can see how the primary stagnation point approaches the secondary one. This can be seen comparing plots of rows (C) and (D) in the first or second columns. 

Returning to the results for  $M_A = 12$ (Figure \ref{fig_M12}), rows (D) and (E)
present results for the large but subsonic values of the ISM Mach number. The heliopause is open in downwind. However, the cross connection bridge between wings of the tube becomes wider for  $M_\infty$=0.5. For $M_\infty$=0.9 it almost disappears and the cross section of the astropause in the plane perpendicular to $x$-direction has elliptical shape. Plot in row E column (5) shows that the distance to the astropause in $y$ direction is twice less than in $z$ direction. Therefore, the effect of stellar magnetic field are still significant.

Rows (F) and (G) present the results of calculations with supersonic values of the ISM Mach number $M_\infty$=1.6 and 2.2, respectively. The structure of the flow pattern that we obtained here is classical supersonic structure with the interstellar bow shock and the Mach disk in the tail (see Figure~\ref{astrosphere_sketches}b). Nevertheless, even in these cases the astropause is elongated toward $z$-axis by about 30 \%, so imprints of the stellar magnetic field on the global shape of the astrosphere are still pronounced.

The Figures \ref{fig_M8},\ref{fig_M4} represent flow patterns and plasma parameter distributions obtained numerically  for lower values of the Alvenic Mach number, $M_A$, i.e. for stronger stellar magnetic field. In general, the flow patterns and their bifurcations with $M_{\infty}$ are similar to those discussed for $M_A =4$, so we will not repeat the discussion. Nevertheless, we observe that for the stronger the stellar magnetic field  the smaller effects of the interstellar flow on the shape of the astropause.
 It is more difficult for the external flow to bend it as it is seen by comparison of the plots in the first two columns in rows (A) and (B) in Figures \ref{fig_M12}-\ref{fig_M4}. The bifurcation of the flow pattern from tube-like to the open astropause appears at $M_{\infty}\approx$0.45 for $M_A$ = 8 and at $M_{\infty}\approx0.95$ for  $M_A$ = 4.

\clearpage
\clearpage
\clearpage

\clearpage

\section{Summary and discussion}

In this paper we study effects of the stellar magnetic field on the global shape of the astrosphere. 
The results can be summarized as follows:

1. The flow pattern of the SW/ISM interaction may have one of three different types depending on the values of two dimensionless parameters which are 1) the gasdynamical Mach number of the interstellar flow, $M_{\infty}$, and 2) Alfven Mach number of the stellar wind, $M_A$.

2. For the small values $M_{\infty}$ the structure of the flow has tube-like structure (Figure \ref{astrosphere_sketches}c) similar to the one that was proposed by \cite{opher15} and studied by  \cite{drake15, golikov2017a, golikov2017b} for the not moving (with respect to the star) interstellar medium. We have shown that the flow of the interstellar gas around the tube-like astropause is quite complex in the tail region. The flow pattern includes the reverse interstellar flow and the stagnation point in the interstellar gas. We call it the secondary stagnation point to distinguish from the primary stagnation point that appears at the astropause.

3. By increasing the interstellar Mach number it was found that for any given $M_A$ there is a critical value of $M_{\text{crit}, \infty}$ at which the flow pattern changes its structure i.e. has bifurcation.  For $M_{\infty}$<$M_{\text{crit}, \infty}$ the follow pattern has a tube-like structure described above. For $M_{\infty}$>$M_{\text{crit}, \infty}$ the  astropause has an open structure in the tail as it is shown in Figure \ref{astrosphere_sketches}a. Although this structure (especially in meridian plane) looks quite similar to the classical global structure of the heliopause in the subsonic stellar wind (e.g. \cite{parker61}),  we found that the astrosphere remains to be very narrow in the equatorial plane (xy-plane). For the Mach numbers slightly higher than $M_{\text{crit}, \infty}$ the astrosphere has a very narrow region in the tail that connects two wings of the tube. The larger $M_{\infty}$ is the wider the size of the astrosphere in the tail is. For Mach numbers close to 1 (e.g. $M_{\infty}=0.9$) the astropause has elliptical shape in the projections to the planes perpendicular to the ISM flow direction.

4. The value of $M_{\text{crit}, \infty}$ increases with the decrease of Alfvenic Mach number that corresponds to the effective increase of the stellar magnetic field.

5. At $M_{\infty} =1$ another bifurcation of the flow pattern appears. The bow shock is formed in the upwind part of the interstellar medium, and the Mach disk is formed in the tail region. The detailed recent description of the 'standard" flow pattern in the tail consisting of Mach disk, reflected shock and secondary tangential discontinuity can be found in \cite{Korolkov_2020}. Note, however, that even for large supersonic Mach numbers $M_{\infty}\approx$ 2-3 the astropause is still elongated toward the axis of the stellar rotation.
For the parameters close to the heliospheric, the astropause is elongated toward the poles by $\approx$30 \%.

In general, we can conclude that the effect of the stellar magnetic field on the interaction region is very important and it was overlooked in the classical works of \cite{parker61}, \cite{baranov70} . 
Theoretically, this influence was considered in the works of \cite{Yu1974} and \cite{Nerney} and it was qualitatively confirmed in our results. The inclusion of the interplanetary field in the global model of the heliosphere is given in the work of \cite{Linde1998}. Our goal was to study the isolated effect of the interplanetary field on the heliopause.
As for the heliosphere, our results demonstrate that the global shape of the heliosphere is open in the tail region and corresponds more to the classical picture than to the tube-like heliopause structure proposed by \cite{opher15}.However, the solar magnetic field influences the global shape of the heliopause. In particularly, the heliopause is elongated toward the poles by about 30 \%.

The model employed in the paper has several simplifications. Some of them are discussed below.
This paper presents the results of 3D modeling of the interaction of the stellar wind with the interstellar medium under the assumption of ideal magneto-hydrodynamics. We neglect the effects of the interstellar magnetic field  (IsMF) and the interstellar atoms of hydrogen (H-atoms). Although we know that both IsMF and H-atoms are extremely important for the global structure of the heliopause (see, e.g. Izmodenov 2000, Izmodenov et al., 2005), we made this simplification to explore the effect of the stellar magnetic field in clear conditions without mixing with other effects. Detailed parametric studies of the joint effects will be performed in the future. However,  the 3D kinetic-MHD model \cite{izmodenov_alexashov2015} that takes into account these effects confirms our conclusions on the open tail heliopause.

In this paper we also neglect the radial component of the stellar magnetic field.
This assumption is justified, since according to the Parker's solution \ref{parker_field} this component
decreases as square of the distance. Therefore, at the termination shock the radial component is much smaller in comparison to the azimuthal component and can be neglected. Nevertheless, the radial component of the magnetic field gives a rise to the azimuthal velocity component of the stellar wind. 
This component is small to have any valuable effects, but it may destroy the symmetry of the flow patterns in (xz) and (xy) -planes that would increase the computational domain by factor of four. Our numerical tests show that taking into account the radial magnetic field does not affect the results presented here.

Note, finally, that in this work we restrict ourselves to the stationary solution of the SW/ISM interaction. In order to obtain the stationary solutions numerically we performed calculations with small value of the parameter $\chi=2$, which is the ratio of the terminal velocity of the stellar wind to the speed of sound of the interstellar medium. We have shown for the stationary solutions that the flow pattern does not depend on $\chi$. Therefore, the obtained solutions give us correct stationary solution for large values of $\chi$. It was pointed out before that for the heliosphere $\chi = 32.234$. It has been shown by \cite{Korolkov_2020} that the numerical solution of the SW/ISM interaction problem in 2D gasdynamic approach becomes unstable. The instability is associated with the Kelvin-Helmholtz instability. We expect that the instability will also lead to unstable time-dependent solution for the problem considered in this paper. The instability will certainly effect the tail structure. We plan to study this problem in the future.

\section{Acknowledgments}
Global heliospheric/astrospheric modeling part of this work has been done in the frame of Russian Science Foundation grant 14-12-01096.

\section{Data availability}
The results of numerical modeling underlying this article will be shared on reasonable request to Sergey Korolkov (korolkovsergey1998@mail.ru).






\bibliographystyle{mnras}
\bibliography{example} 

\bsp	
\label{lastpage}

\end{document}